\documentclass{JHEP3}

\usepackage{amsmath, amssymb}
\usepackage{graphicx}

\newcommand{\mathsym}[1]{{}}

\renewcommand\({\left(}
\renewcommand\){\right)}
\renewcommand\[{\left[}
\renewcommand\]{\right]}

\newcommand{\dd}{{\rm d}}
\newcommand{\e}{{\rm e}}

\newcommand\eps{\epsilon}

\def\ba{\begin{eqnarray}}
\def\ea{\end{eqnarray}}
\def\be{\begin{equation}}
\def\ee{\end{equation}}

\def\L{\mathcal{L}}

\def\nn{\nonumber}
\def\({\left(}
\def\){\right)}

\newcommand{\roughly}[1]{\mathrel{\raise.3ex\hbox{$#1$\kern-0.85em
\lower1ex\hbox{$\sim$}}}}

\DeclareMathOperator{\diag}{diag}

\title{Avoiding the dangers of a soft-wall singularity}

\date{\today}

\preprint{NIKHEF/2011-016}

\author{Damien P. George and Marieke Postma\\
Nikhef Theory Group,\\
Science Park 105, 1098 XG Amsterdam, The Netherlands\\
\vspace{-5pt}\\
E-mail: \email{dpgeorge@nikhef.nl}, \email{mpostma@nikhef.nl}}

\abstract{We critically analyse the nature of the infrared
singularity in Randall-Sundrum soft-wall models, where the extra
dimension is dynamically compactified by the formation of a curvature
singularity.  Due to the Israel junction conditions, this singularity
can only be shielded by a time-independent black-hole horizon if there
is ghost matter on the UV brane.  For this construction the spectrum
of 4D states is shown to be similar to the original soft-wall case.
We point out, however, that no such shielding is needed, as the
singularity satisfies unitary boundary conditions.}


\begin{document}


\section{Introduction}

Soft-wall models are a recent modification to the Randall-Sundrum (RS)
scenario, where a compact extra dimension is warped in order to solve
the electroweak hierarchy problem~\cite{Randall:1999ee}.  In soft-wall
models, the warp factor exponent diverges in the infrared (IR) and the
negative tension IR brane is replaced by a curvature singularity.  The
original motivation for such warping was to obtain Kaluza-Klein (KK)
modes with mass squared that is linear in KK number, with the aim to
better model excited mesons in QCD using the AdS/CFT
correspondence~\cite{Karch:2006pv}.  Since this initial work, soft
walls have been used to construct extra-dimensional extensions of the
standard model~\cite{Batell:2008zm, Falkowski:2008fz,Batell:2008me,
Delgado:2009xb,MertAybat:2009mk,Gherghetta:2009qs,Cabrer:2009we,vonGersdorff:2010ht,Gherghetta:2010he},
as well as a holographic dual description of unparticle
models~\cite{Cacciapaglia:2008ns, Falkowski:2008yr}.  One of the main
motivations of a soft-wall background is that it allows a lower KK
energy scale without violating electroweak observables which are
problematic in the original RS
framework~\cite{Atkins:2010cc,Cabrer:2010si,Cabrer:2011fb}.

The minimal soft-wall spacetime is supported by a dilaton field which
also diverges in the IR, and the distance to the singularity is
stabilised by an ultraviolet (UV) localised potential for the
dilaton~\cite{Cabrer:2009we,Gherghetta:2010he}.  Alternatively, the UV
brane can be replaced by a domain-wall and the set-up stabilised by
parity and an appropriate bulk
potential~\cite{Aybat:2010sn,George:2011tn}.  With suitable scalar
potentials, soft-wall models can solve the hierarchy problem in the
same way as in RS~\cite{Cabrer:2009we}.  A basic feature of these
models which remains to be analysed in detail is the nature of the
singularity itself, and this shall be the focus of the current paper.
The soft-wall singularity is allowed if it does not lead to
pathological behaviour, such as loss of unitarity or instabilities.

Cosmic censorship conjectures that all naked singularities (such as
those in the soft-wall set-up) are protected by an event horizon.  In
the soft-wall literature to date, it has been assumed that such a
horizon can be constructed to shield the bulk from the singularity.
This assumption is based on the early work of
Gubser~\cite{Gubser:2000nd}, where it is shown that a physical
spacetime can be constructed with a singularity and an event horizon,
provided the scalar potential is bounded above in the solution. The
Hawking temperature of the horizon can be identified with the finite
temperature in the dual field theory which serves as an IR
cutoff. Spacetime is asymptotically AdS in a direction away from the
horizon.  To apply this result to the soft-wall scenario requires
patching two truncated versions of the Gubser spacetime back-to-back,
so that the singularities bound the (almost) AdS interior.  We shall
show in this paper that such a patching cannot be accomplished without
putting ghost matter at the junction.  These results follow simply
from the inability to match the Israel junction conditions when the 5D
metric is generalised to allow for a horizon.  In such a case, the
relevant Einstein's equations require ghost matter to bend the metric
components in the right way.

Having ghost matter on the UV brane is particularly unsettling. An
obvious way around this is to look for background configurations with
a horizon and no ghosts, but which are time dependent in order to
satisfy the Israel junction conditions and Einstein's equations.  If
such solutions have a life-time longer than the age of the universe,
they provide a viable background.  It is however questionable whether
this can be achieved. If the soft-wall model is to solve the hierarchy
problem, a much larger decay rate of the order of the electroweak
scale seems natural (as, besides the Planck scale, there is no other
scale in the system).

Thus shielding the soft wall singularity by a horizon cannot be
achieved in a straightforward manner. As we shall point out in this paper,
there is actually no need to do so.  The soft-wall singularity is
actually not ``visible'', and one can thus argue that cosmic censorship
does not apply.  Even though the singularity is at a finite distance in
physical coordinates, there are no plane wave modes in the spectrum
that can travel to it in a finite time.  The KK spectrum
consists of discrete modes that all have an extra-dimensional profile
that dies off rapidly towards the singularity.\footnote{In the special
case that the spectrum consists of unparticles, that is, the spectrum
is continuous but with a mass gap, the continuous modes still die off
rapidly towards the singularity.}  As a consequence, the system
satisfies unitary boundary conditions, which guarantee that no
conserved charge, such as energy and (angular) momentum, can leak into
the singularity ~\cite{GellMann:1985if, Cohen:1999ia, Gremm:2000dj}.
Without this leakage there are no pathologies, and unitarity is
automatically conserved.

The paper is organised as follows.  In Section~\ref{sec:sw} we start
with a derivation of the classical background solution. We generalise
the metric of the soft-wall set-up to allow for a horizon shielding
the singularity; we call this the ``black-wall'' configuration. The
rest of this section is devoted to a discussion the of the original
soft-wall model with a naked singularity.  We also briefly comment on
the spectrum of KK states. In Section~\ref{sec:bw} we consider the
problems that arise if one tries to shield the singularity by a
black wall, followed in Section~\ref{sec:bwpert} by a discussion of
the perturbative spectrum of a black wall.  In Section~\ref{sec:BC}
we argue that the soft-wall singularity is not ``visible'', as the
KK modes all satisfy unitary boundary conditions.  Hence, no
shielding by a horizon is needed.  We conclude in
Section~\ref{sec:concl}.


\section{The soft-wall model}
\label{sec:sw}

In this section we give a concise review of the soft wall-model,
focusing on those features that later enter the discussion on the
nature of the singularity.  We begin by deriving the equations of
motion and boundary conditions for a metric that is general enough to
include an event horizon shielding the singularity.  From these
equations one can obtain the original soft-wall background solution,
to be discussed in the remainder of this section, as well as the
black-wall solution discussed in the following section.

\subsection{The background equations}

There are three main ingredients for a soft-wall model: something to
warp the metric at the origin (the UV brane or domain wall), something
to diverge the metric at the edge (the dilaton), and some
stabilisation mechanism (brane potential terms).  In the minimal
set-up we discuss here, a UV brane with a dilaton suffices.
Stabilisation is ensured by a brane-potential for the dilaton, fixing
its BC there; see~\cite{Cabrer:2009we}.  The nature of the singularity
only depends on the IR details of the model, in particular on the
asymptotic behaviour of the bulk potential $V$ with respect to the
field playing the role of the dilaton.

Consider then 5D spacetime with gravity, a single real scalar field
$\phi$ in the bulk (the dilaton), and a fundamental 3-brane with
localised matter placed at $y_\text{br}=0$.  Using normal Gaussian
coordinates, the most general static metric is of the
form~\cite{Gubser:2000nd}\footnote{For $f(y)$ not constant 4D Lorentz
invariance is broken if bulk fields have different 5D profiles.
For $f$ nearly constant, except close to the horizon/singularity,
these effects can presumably be made arbitrarily small.
See~\cite{George:2008vu} for further details of such Lorentz
violation.}
\be 
\dd s^2 = \e^{-2\sigma(y)} (- f(y)\dd t^2 + \dd \vec{x}^2) + 
  \frac{\dd y^2}{f(y)} \:.
\label{eq:metric}
\ee
We could redefine $y$ to absorb $f$ in the $\dd y^2$ term, but the
current form of the metric allows us to clearly see the formation of
an event horizon.  Indeed, this occurs at $y_h$ where $f(y_h)=0$.
Since all metric functions depend only on $y$, we can rescale all
coordinates by a constant to set $f(0)=1$ and $\sigma(0)=0$.  The 5D
action consists of a bulk and brane contribution:
\begin{equation}
\begin{aligned}
S =
  & \int \dd^5 x \sqrt{-g_5} \[
    \frac{R}{6\kappa^2}
    - \frac{1}{2} g_5^{MN} (\partial_M \phi)(\partial_N \phi)
    - V(\phi)
  \] \\
  & + \int \dd^5 x \sqrt{-g_\text{br}} \[
    -\lambda(\phi)
    + \L_\text{br}(\psi, X)
  \] \delta(y-y_\text{br}) \:,
\label{eq:S}
\end{aligned}
\end{equation}
where $(g_5)_{MN}$ is the 5D metric and $\kappa^2$ is proportional
to the 5D Newton's constant.  The induced metric on the brane is
defined via
\be
(g_\text{br})_{\mu\nu} =
  \frac{\partial x^M}{\partial z^\mu}
  \frac{\partial x^N}{\partial z^\nu} 
  (g_5)_{MN} = \delta^M_\mu \delta^N_\nu (g_5)_{MN} \:,
\ee
where $z^\mu$ are the coordinates on the brane and $x^M$ the
coordinates in the bulk.  The bulk scalar $\phi(x^\mu,y)$ has a
brane-localised potential $\lambda(\phi)$, and we allow for
additional brane matter $\psi(x^\mu)$ through the Lagrangian
$\L_\text{br}$.  This Lagrangian is generically a function of the
kinetic term
$X = -(1/2) g_\text{br}^{\mu\nu} \partial_\mu \psi \partial_\nu \psi$;
for canonical kinetic terms the brane Lagrangian is
$\L_\text{br} = X - V_\text{br}(\psi)$.  

The energy momentum tensor for the action~\eqref{eq:S} is
$T_{MN}= (-2/\sqrt{-g_5}) \partial S/\partial g_5^{MN}$, and takes
the explicit form
\begin{equation}
\begin{aligned}
T_{MN} =
  \: & (g_5)_{MN} \(
    -\frac12 g_5^{PQ} (\partial_P \phi)(\partial_Q \phi)
    - V(\phi)
  \)
  + (\partial_M \phi)(\partial_N \phi) \\
  & + \frac{\sqrt{-g_\text{br}}}{\sqrt{-g_5}}
    \delta_M^\mu \delta_N^\nu \delta(y-y_\text{br}) \[
      (g_\text{br})_{\mu\nu} (-\lambda(\phi) + \L_\text{br})
      + \frac{\partial \L_\text{br}}{\partial X} (\partial_\mu \psi)(\partial_\nu \psi)
    \] \:.
\end{aligned}
\label{eq:T}
\end{equation}
Splitting the energy-momentum tensor into pieces that do and do not
depend on the brane matter field $\psi$, we can identify the piece
that does depend on $\psi$ as a cosmological fluid localised to
the brane, with $T^M_N=\delta(y-y_\text{br}) \diag(-\rho_\text{br},
p_\text{br},p_\text{br},p_\text{br},0)$.  The brane energy and
pressure components then read
\begin{align}
  \rho_\text{br} &= -\L_\text{br}
    + \frac{\partial \L_\text{br}}{\partial X} \dot{\psi}^2
  = \frac12 \dot{\psi}^2 + V_\text{br} \:,
\label{rhobr} \\
  p_\text{br} &= \L_\text{br}
  \hspace{59pt} = \frac12 \dot{\psi}^2 - V_\text{br}  \:.
\label{pbr}
\end{align}
We have used $f(0)=e^{\sigma(0)}=1$, and also the off-diagonal
Einstein's equations to obtain $\nabla \psi=0$.
The last equality in the above definitions is only for canonically
normalised brane kinetic terms such that
$\partial \L_\text{br}/\partial X=1$.
We shall express all $\psi$ quantities in terms of $\rho_\text{br}$
and $p_\text{br}$, and treat it as an arbitrary cosmological
source localised to the brane.  When we need to study the form
of the energy and pressure in more detail (such as the condition
for ghosts) we shall use the above definitions.

The equations of motion for the system are Einstein's equations
$G^M_N = 3 \kappa^2 T^M_N$, and the Euler-Lagrange equations.
The bulk off-diagonal Einstein's equations enforce $\partial_\mu\phi=0$,
a consequence of our diagonal and isotropic metric ansatz.
Then, the Euler-Lagrange equation and the $(00)-(ii)$, $(00)-(55)$
and $(55)$ Einstein's equations read:
\begin{align}
f \phi'' - 4 \sigma' f \phi' + f' \phi' - \partial_\phi V &=
  \delta(y-y_\text{br}) \partial_\phi \lambda \:,
\label{eq:EL}\\
f'' - 4f'\sigma' &= 6\kappa^2 \delta(y-y_\text{br}) (\rho_{\rm br} + p_{\rm br}) \:,
\label{eq:ee0i} \\
f(\sigma'' - \kappa^2{\phi'}^2) &
= \kappa^2 \delta(y-y_\text{br}) (\lambda + \rho_{\rm br}) \:,
\label{eq:ee05} \\
4 f \sigma'^2 - f' \sigma' - \kappa^2 (f \phi'^2 - 2 V) &= 0 \:.
\label{eq:eeconst}
\end{align}
Only three of these four equations are independent. In addition there
is the brane localised Euler-Lagrange equation for the brane matter
$\psi$, which will not be important for our purposes.

The above system has five independent integration constants.
As mentioned before, $f(0) = 1$, $\sigma(0)=0$ can be fixed by
reparameterisation of the coordinates.  The remaining three are
$\{f'(0_+),\sigma'(0_+),\phi'(0_+)\}$.  Once these constants are
known, $\phi(0)$ is fixed by the constraint equation
\eqref{eq:eeconst} evaluated at $y=0_+$ [the values of all functions
must be continuous over the brane, so $\phi(0)=\phi(0_\pm)$].
The three derivative integration constants are all fixed by the
boundary conditions, that is, by the matter on the brane.
Integrating the above equations over the brane, and assuming a
$Z_2$-symmetry, these boundary conditions are
\be
f'(0_+) = 3\kappa^2(\rho_\text{br} + p_\text{br}) \:,
\qquad
\sigma'(0_+) = \frac12 \kappa^2 (\lambda + \rho_\text{br}) \:,
\qquad
\phi'(0_+) = \frac12 \partial_\phi \lambda \:.
\label{eq:BC}
\ee
Given a choice of matter content on the brane, and brane potential
terms, the model is stabilised precisely because there are no choices
for the integration constants.

One important consistency check on the theory is that the
effective 4D action has a vanishing cosmological constant.  This
is because 4D slices of our metric ansatz are time-independent.
The 4D action is obtained by integrating the 5D action~\eqref{eq:S}
over the extra dimension.  The 5D Ricci scalar is
$R = 9 f' \sigma'-20 f \sigma'^2 - f''+8 f \sigma''$,
and $\sqrt{-g_5} = \e^{-4\sigma}$.  Use the equations of motion
(\ref{eq:ee0i}-\ref{eq:eeconst}) to eliminate $V$, $f''$ and $\phi'$.
The result, including the brane term, is then
\begin{align}
S_5 &= \int \dd^4 x \int \dd y \left\{
  \frac{1}{6\kappa^2} \e^{-4 \sigma} \(2f' \sigma' -8f \sigma'^2+2f \sigma'' \)
  + \delta(y-y_\text{br}) \(-p_\text{br} + \L_\text{br} \) \right\}
\nn \\
&= \int \dd^4 x \left\{ \frac{1}{3\kappa^2} \[ f \sigma' \e^{-4 \sigma} \]^{y_s}_{-y_s}
\right \} \:.
\label{eq:surface}
\end{align}
To get flat 4D Minkowski spacetime this boundary term should vanish,
and its vanishing depends on the model.

\subsection{The soft-wall solution}

In the usual soft-wall set-up the background solution has
$f(y)=1$ (see for example~\cite{Cabrer:2009we}).  The boundary
conditions can be satisfied in this case by the choice of
vanishing brane matter, $\rho_\text{br}=p_\text{br}=0$ (a
cosmological constant $\rho_\text{br}+p_\text{br}=0$ is already
accounted for by $\lambda$).  Having a constant $f$ and no
brane matter means equation~\eqref{eq:ee0i} is satisfied
identically, and 4D slices of the metric at constant $y$ are
Poincar\'e.  The slope $\sigma'(0_+)$ is positive for a positive
tension brane ($\lambda >0$) and energy scales are warped down
towards the IR as one moves out into the bulk.  In this case the
model can solve the electroweak hierarchy problem as detailed
in~\cite{Cabrer:2009we}.

To see the formation of a singularity in the IR region, and a
corresponding dynamical truncation of the extra dimension, we must
solve the background equations of motion.  In Ref.~\cite{Cabrer:2009we}
the fake supergravity formalism~\cite{DeWolfe:1999cp, Freedman:2003ax}
is used to rewrite the equations of motion in first order form,
allowing the construction of an explicit model.  This approach
requires the introduction of superpotential $W(\phi)$ such that
the full potential can be written as
$V = (1/2) (\partial_\phi W)^2 - 2\kappa^2 W^2$.
Then the bulk equations of motion are satisfied if
\be
  \sigma' = \kappa^2 W \:,
  \qquad \phi' = \partial_\phi W \:,
\ee
while the boundary conditions read
\be
  W(\phi_0) =\frac12 \lambda(\phi_0) \:, \qquad
  \partial_\phi W(\phi_0) = \frac12 \partial_\phi \lambda(\phi_0) \:,
\ee
where $\phi_0=\phi(0)$.
Consider then the class of superpotentials with asymptotic
behaviour
\be
  \lim_{\phi \to \infty} W = c (\kappa\phi)^\beta \e^{\nu\kappa\phi} \:,
  \qquad (\nu >0) \:.
  \label{eq:limW}
\ee
We take $\nu \neq 0$ as polynomial functions cannot solve the
hierarchy problem.  The field $\phi$ (and consequently also the
metric function $\sigma$ and the Ricci scalar) blows up at
\be
y_s = \int_{\phi_0}^\infty \frac{\dd \phi}{\partial_\phi W} \:,
\ee
which occurs at a finite and positive value if $\nu>0$ and $c>0$
(the equations of motion are solved on the interval $0<y<y_s$).
The extra dimension is cut off dynamically by a singularity.
This is the soft-wall.

The solution gives rise to a 4D Minkowski spacetime provided the
surface terms \eqref{eq:surface} vanish for $f(y)=1$.  At the
singularity, $y \to y_s$, the field $\phi \to \infty$, and the
boundary term only depends on the asymptotic form of the
superpotential, which we parameterise as before \eqref{eq:limW}.  It
follows that in this limit $\sigma \to \kappa\phi/\nu -
(\beta/\nu^2)\ln(\nu\kappa\phi)$, and thus
\be
\lim_{y\to y_s} \sigma' \e^{-4 \sigma}
  \propto (\kappa\phi)^{\beta(1+4/\nu^2)} \;
  \e^{\nu\kappa\phi(1 - 4/\nu^2)} \to 0
\quad \text{iff} \quad \nu <2 \:.
\ee
Furthermore, in order to have a phenomenologically viable model with a
mass gap in the spectrum of KK states, one requires $\nu\ge1$.
Therefore, a viable soft-wall model that solves hierarchy problem is
only possible for $1 \leq \nu < 2$~\cite{Cabrer:2009we}.

\subsection{The Kaluza-Klein spectrum}
\label{sec:KK}

For the original soft-wall model, we parameterise the complete set
of metric perturbations as~\cite{Csaki:2000zn}
\be
\dd s^2 = \e^{-2\sigma} 
(1-2F) \[-\dd t^2 + (\delta_{ij} + h_{ij}) \dd x^i \dd x^j\]
+ (1+4F) \dd y^2
\ee
and the dilaton perturbation as $\phi(x,y) = \phi(y) + \delta
\phi(x,y)$.  $F=F(x,y)$ and $h_{ij}(x,y)$ is transverse and
traceless.  As before
$\sigma(y),\, \phi(y)$ are the background solutions.  The scalar
($\delta \phi, F$) and tensor perturbations ($h_{ij}$) decouple at
first order and can be analysed separately. Einstein's equations yield
a constraint equations for the scalar perturbations
\be
\kappa^2 \phi' \delta \phi= F'-2\sigma' F \:,
\label{eq:constraintF}
\ee
leaving one scalar degree of freedom.  There are two tensor degrees
of freedom corresponding to the two polarisations of the graviton.
We perform separation of variables for the scalar perturbations and
tensor perturbations:
\be  F(x^\mu, y) = F(y) \rho(x^\mu) \:, \quad \delta \phi
(x^\mu, y) = \delta \phi (y) \rho(x^\mu) \:, \quad
h_{ij}(x^\mu,y) = h(y) \eps_{ij}(x^\mu) \:.
\label{eq:separation}
\ee
The 4D perturbations satisfy $\Box \rho = m^2 \rho$ and $\Box
\eps_{ij} = m^2 \eps_{ij}$.  It is now convenient to go to conformally
flat coordinates ($\dd y = \e^{-\sigma} \dd z$), and rescale the
perturbations, to obtain a Schr\"odinger equation for the profile
functions
\be
-\partial_z^2 \tilde q + V_q \tilde q = m^2 \tilde q
\ee
with $\tilde q = \{\tilde F,\,\tilde h\} = \e^{-3\sigma/2}\{
F/(\partial_z \phi),\, h\}$.  This equation is solved to find the
KK spectrum of modes, including the profile functions.  Substituting
these solutions into the original 5D action, expanding to second order,
and integrating out the extra dimension yields the effective 4D action.
This procedure requires partial integration, and surface terms are
obtained.  Demanding that these surface terms vanish, so to obtain 4D
Minkowski space, gives the boundary equations for the tensor mode
\be
\e^{-4\sigma} h h' |_{y_s} = 0 \:.
\label{eq:surface1}
\ee
For the scalar perturbation there is a boundary condition first order
in the perturbations as well (see \cite{George:2011tn} for a
derivation)
\be
\begin{aligned}
\e^{-4\sigma} (F'-6\sigma' F) |_{y_s} &= 0 \:,\\
\e^{-4\sigma} \(\frac13 FF' - 10 \sigma' F^2 
+\frac{\kappa^2}2 \delta\phi \delta \phi' \) \big|_{y_s}&= 0 \:.
\end{aligned}
\label{eq:surface2}
\ee
These are the first and second order extensions of the zeroth order
result \eqref{eq:surface}.  Using the constraint equation
\eqref{eq:constraintF}, and the asymptotic behaviour of the background
near the singularity,\footnote{Near the singularity we have
$\sigma\sim(-1/\nu^2)\ln k(y_s-y)$ and
$\kappa\phi\sim(-1/\nu)\ln k\nu^2(y_s-y)$.}
we find that all first and second order terms above (such as
$\e^{-4\sigma}FF'$) should vanish separately.  In essence, there
can be no cancellations of badly-divergent behaviour among the
perturbations $F$ and $\delta\phi$ in the two expressions above,
at least in the parameter range of interest, $1\le\nu<2$.
In conformally flat coordinates the boundary
condition for, for example, the tensor perturbation translates to
\be 
\e^{-3\sigma}h \partial_z h|_{z_0,z_s}=
\tilde h ( \partial_z \tilde h + \frac32 \partial_z \sigma \tilde h)
|_{z_0,z_s} = 0
\ee
In addition the mode functions should be normalisable:
$\int \tilde q^2 \dd z < \infty$.

The potential $V_q$ depends on the background solutions, and is given
explicitly in \cite{Cabrer:2009we}.  The spectrum is discrete with a
mass gap for $\nu >1$ and continuous with a mass gap for the boundary
value $\nu =1$.  The mass gap is of order $m \sim
k(ky_s)^{-1/\nu^2}\e^{-k y_s}$, with $k$ of the order of the 5D Planck
mass.  This yields electroweak scale masses for $ky_s \sim 30$, just
as in RS.


\section{The black-wall solution}
\label{sec:bw}

The soft-wall solution has a naked singularity at finite coordinate
distance, so matter can potentially reach it in a finite time.
This may lead to problems with loss of unitarity or unknown quantum
gravity corrections.  The usual assumption in the soft-wall literature
to date is that the singularity can be shielded by a black-wall
horizon, which builds on earlier work by Gubser \cite{Gubser:2000nd}.
This is in line with the cosmic censorship conjecture.  The black-wall
solution corresponds to a non-trivial function $f$ in the metric
\eqref{eq:metric} which becomes zero at some finite coordinate
distance $f(y_h)=0$, with $y_h < y_s$ the position of the black-wall
horizon.  At the horizon there is infinite time dilation, and it takes
matter an infinite time to reach it (as seen by an asymptotic
observer).
The singularity at $y=y_s$ is therefore
shielded by this event horizon.

By way of background, we first discuss the construction of
Gubser~\cite{Gubser:2000nd}.  There, singular spacetimes were
considered arising in theories with bulk matter, and which were
asymptotically AdS in one half of the extra dimension.  This
corresponds to our set-up (with $k>0$) if the boundary conditions
are modified: the brane at the origin is removed, and one half
of the space is AdS, while the other half ends at a singularity
at $y=y_s$.  The bulk solutions to the equations of motion are the
same as in our case, but there are no longer additional boundary
conditions to satisfy at the brane position.  Asymptotically, as
$y \to -\infty$, spacetime approaches pure AdS, $\sigma' \to k$ a
constant, and $f' \to 0$.  In~\cite{Gubser:2000nd} it is shown
that provided the potential $V$ is bounded from above in the
solution, a family of black hole solutions shielding the
singularity exist, suggesting cosmic censorship is at work in such
a spacetime.

However, these conclusions do not have a straightforward application
to the soft-wall model with a brane and $Z_2$ orbifold symmetry, as
now there are also the boundary conditions at the brane position to
satisfy~\eqref{eq:BC}.  The equations of motion~\eqref{eq:ee0i}
and~\eqref{eq:ee05} can be solved in the bulk by
quadrature~\cite{Gubser:2000nd}:
\begin{align}
\sigma &= k y + \kappa^2 \int_0^y \dd y_1 \int_0^{y_1} \dd y_2 \, \phi'(y_2)^2 \:,
\label{eq:solsigma}
\\
f &= 1 + A |k| \int_0^y \dd y_1 \, \e^{4\sigma(y_1)} \:,
\label{eq:solh}
\end{align}
normalised such that $\sigma(0) \equiv \sigma_0 = 0$ and
$f(0) \equiv f_0 = 1$, and with $\sigma_0' = k$ and $f_0'= A |k|$.
The dimensionless constant $A$ controls the location of the
horizon of the black wall.
Except close to the singularity, spacetime is near AdS with
$\sigma \approx k y$.  Only near to the singularity does the
second term in~\eqref{eq:solsigma} become important.  Note this
term is always positive and $\lim_{\phi \to \infty} \sigma = +\infty$.
For $k<0$ this implies the sign of $\sigma'$ flips.  (For
completeness, in the following discussion we consider both
positive and negative $k$.)

To form a horizon requires $A<0$.  Then $f$ is guaranteed to
pass through zero at some point before the singularity, since
$\sigma\to\infty$.  Thus $A<0$ is a necessary and sufficient condition
to achieve a black-wall solution.  If the horizon is close to the
singularity, then in the whole bulk region the soft-wall ($f=1$) and
black-wall ($f\ne1$) solutions are nearly the same, and one might
expect that the soft-wall spectrum analysis carries over to the
black-wall case. This is confirmed by our analysis of the perturbative
spectrum in Section~\ref{sec:bwpert}.  This is a desirable scenario, as
then one retains all the merits of the soft-wall set-up.  For $k<0$
and $|A| \lesssim 1/|k|y_s$ one always has $y_h \approx y_s$, as the second
term in \eqref{eq:solh} is only important near the singularity. On the
other hand, for the usual case of $k>0$ the black-wall horizon is close
to the singularity only for very small $|A| \lesssim e^{-4ky_s}/ky_s$.
Recall the KK scale is $m \sim k \e^{-ky_s}$ with $k$ of the
order of the 5D Planck scale, so the bound on $A$ gives
\be
|A| \lesssim \left( \frac{m}{k} \right)^4
  \left( \ln \frac{k}{m} \right)^{-1}
  \sim 10^{-62} \:,
\ee
where we have used $m=1\;\text{TeV}$ and $k=10^{18}\;\text{GeV}$.

The black-wall solution requires $A|k|=f'_0<0$, which, in combination
with the boundary conditions~\eqref{eq:BC}, implies
$\rho_\text{br}+p_\text{br} < 0$.  This brakes the null energy
condition, implying some kind of ``ghost matter'' is needed on the
brane.  Indeed, for the case of brane-localised scalar matter this
implies negative kinetic terms:
\be
\rho_\text{br}+p_\text{br}
  = \frac{\partial \L_\text{br}}{\partial X} \dot \psi^2
  < 0 \:.
\ee
Thus, even though the bulk solution is static, the matter on the
brane is not --- though it may still be stationary.  The second
boundary condition in~\eqref{eq:BC} gives $(\lambda+\rho_\text{br})>(<)0$
for $k>(<)0$, which can be easily satisfied, irrespective of
$\rho_\text{br}$, by introducing a suitably large positive
(negative) value for $\lambda$; this is effectively the brane
tension.

To get an order of magnitude for the required ghost matter density,
use the bound on $A$ above (for the case $k>0$) to get
\be
\rho_\text{br}\lesssim m^4 \left( \frac{M_*}{k} \right)^3
  \left( \ln \frac{k}{m} \right)^{-1}
  \sim m^4 \:,
\ee
where $M_*=\kappa^{-3/2}$ is the fundamental 5D Planck scale;
$M_*\sim10^{19}\;\text{GeV}$.  Thus, even for a tiny value of
$|A|\sim10^{-62}$ we require a relatively large density
$\rho_\text{br}\sim(1\;\text{TeV})^4$.
For cosmological constant-like values of
$\rho_\text{br}\sim(10^{-11}\;\text{GeV})^4$, one obtains an
extremely small value for $A$: $|A|\sim10^{-119}$.  Note that such
a small value still generates a black-wall solution.

To summarise, the bulk solution for $f$ is fixed by
equation~\eqref{eq:solh} and allows $f$ to pass through zero, and
hence form an event horizon, if and only if $A<0$.  In the soft-wall
set-up, one must patch together two spacetimes with a singularity (or
have reflective boundary conditions), and the resulting Israel
junction conditions can only be satisfied when ghost matter is present
on the central brane.  It is not clear if ghosts, or
phantoms~\cite{Caldwell:1999ew}, can be accommodated in a theory
without instabilities, and without violating cosmological (and other)
constraints~\cite{Cline:2003gs}.

Is there anyway around this conclusion?  One could try to go from 4D
Minkowski to some cosmological time-dependent spacetime.  But this
will not affect the boundary conditions, as they only involve
discontinuities across the brane [that is, in the $y$-coordinate;
see equation~\eqref{eq:nab-bc}].  Adding
extra bulk or brane matter will also not change the situation; this
can immediately be seen from the fact that the boundary conditions are
phrased in terms of total brane energy and pressure, and only depend
on the statement that $A<0$.  Note in this respect that the solution
\eqref{eq:solh} depends purely on the metric, with the bulk matter
entering only indirectly.  From this solution for $f$ we can see that
there is no way to have $f'=0$ at some location and then have $f'\ne0$
at some other location.  Similarly we cannot have $f'$ change sign
in the bulk.

We have so far assumed a $Z_2$ symmetry across the brane, but
this also is not an essential ingredient.  As long as there are
discontinuities along the extra dimension, boundary conditions
exists that can only be satisfied by violating the null energy
condition.  An analogy can be made here with the warp factor in a
compact RS set up.  The relevant equation is
$\sigma''=\kappa^2\phi'^2+\delta(y-y_\text{br}) \lambda$, which
shows that positive energy matter can only bend $\sigma$ upwards.
To construct a compact space, $\sigma$ must bend down at some
point in order to repeat the solution.  This requires a negative
tension brane, $\lambda<0$, at the IR fixed point.  Similarly, a
non-trivial $f$ must always bend either up or down ($A>0$ or $A<0$
respectively), and the direction of bending can only be changed by
matter with $\rho+p<0$.

The above concerns static bulk solutions.  It may still be possible
that there exist stationary or long-lived black-wall solutions.  For
example, consider placing two Schwarzschild black holes in the same
spacetime.  This cannot be done with a static metric unless there
is some exotic matter between the black holes which shield their
gravitational effects from one another.  One solution is to put them
in orbit around each other.  Or, if they are far enough apart, they
will only interact very weakly and constitute a very long-lived
solution.  Either way, one requires time dependence in the metric to
find physical solutions.  Similarly, two black walls will gravitate
unless one arranges the precise form of ghost matter (which behaves
like anti-gravity) to shield them from one another.

In analogy with the Schwarzschild black holes, it may be possible
to construct a black-wall solution that does not require ghosts, but
instead uses a more general, time-dependent metric. The assumption
here is that any useful time-dependent solution closely resembles the
static black-wall configuration.  Finding such a solution is
difficult due to the non-linear nature of the equations, and is
beyond the scope of this paper.  But it would be interesting to see
whether such a solution can be constructed, and whether it has a
cosmologically long decay time, making it phenomenologically viable.
Besides the Planck scale, the electroweak scale is the only other
scale in the problem, and one might naively think that it sets the
decay rate of the system, leading to rapidly evolving black walls.


\section{Perturbations around the black-wall background}
\label{sec:bwpert}

In this section we develop the equations governing the scalar
perturbations in the black-wall background.  Since in the presence
of a black-wall four dimensional Lorentz symmetry is broken, this
is a non-trivial generalisation of the soft-wall perturbations
discussed in Section~\ref{sec:KK}.  We do not concern ourselves
here with the tensor perturbation $h_{ij}$.  In the transverse
traceless gauge these degrees of freedom appear only with the
spatial coordinates $\vec{x}$ and the introduction of the
black-wall metric factor $f$ will not alter the components of
$h_{ij}$.  Nevertheless, the spin-2 spectrum will be modified, but
for small $A$ we expect it to be a minimal change, and the change
should follow qualitatively what happens in the spin-0 sector.

If one is willing to admit ghost matter in the theory then the
black-wall solution is valid and the singularity is shielded by a
horizon.  In this case the classical stability of such a
configuration as a solution to Einstein's equations must be checked,
which requires looking at perturbations of the set up.  As per the
discussion following~\eqref{eq:BC}, since there are no free
integration constants in the solution, there is no zero-mode in
the spectrum.  If all KK modes have positive mass, the solution is
classically stable.  In addition to checking the stability, it is
also interesting to see whether the spectrum of KK modes is
qualitatively the same as in the soft-wall background. If so, the
soft-wall solution to the hierarchy problem straightforwardly
carries over to the black-wall case.

Consider the metric
\be
\dd s^2 = - n^2(t,y) \dd t^2 + a^2(t,y) \dd \vec{x}^2 + b^2(t,y) \dd y^2 \:.
\label{eq:nab-metric}
\ee
For comparison with our previous ansatz~\eqref{eq:metric}:
$n^2 = f \e^{-2 \sigma}$, $a^2 = \e^{-2\sigma}$ and $b^2 = 1/f$.
Using the same action as before, Einstein's equations yield the
Israel junction condition
\be
\[\frac{n'}{n} - \frac{a'}{a}\]_{y_\text{br}} = 3 \kappa^2 b (\rho + p) \:,
\label{eq:nab-bc}
\ee
where the jump operator is $[X]_y=\lim_{\eps\to0} [X(y+\eps)-X(y-\eps)]$.
Fix the gauge by choosing $n(0)=a(0)=1$.  If $n'(0)<a'(0)$ (which
is what we want for a black-wall solution) then we must have ghost
matter.  If we instead choose $n'(0)=a'(0)$ (or $n'(0)>a'(0)$)
then we do not need ghosts, but instead need the time-dependence
in the metric factors to turn $n$ over in the bulk so that it
passes through zero at some horizon value $y_h$.  Determining
whether or not this is possible involves solving Einstein's
equations with the general metric ansatz~\eqref{eq:nab-metric},
something which we do not attempt here.  Instead, we assume ghost
matter and proceed to compute the KK spectrum.

The scalar perturbations to the metric~\eqref{eq:nab-metric}
are~\cite{Bridgman:2001mc,Deffayet:2002fn}
\be
g_{AB} =
\begin{pmatrix}
-n^2(1-2F) & n a B|_i & nbJ \\
n a B|_i & a^2[(1-2G)\delta_{ij} + 2E|_{ij}] & ab B_y|_i \\
nbJ & ab B_y|_i & b^2(1+ 2H)
\end{pmatrix} \:.
\label{eq:deltaG}
\ee
A vertical bar denotes a derivative with respect to $x^i$.
Using a gauge transformation we can set $E=B=B_y=0$ (three degrees
of freedom by choosing $\{\delta t,\,\delta x,\,\delta y\}$).
With $f'\ne0$, Einstein's equations require $J\ne0$, and one
cannot make a clear identification of a physical mode in the usual
sense (that is, a combination of the degrees of freedom that
satisfies a Schr\"odinger-like equation).  This is in part due to
the fact that separation of variables between 4D and $y$ does not
go through, since different 4D slices at constant $y$ of the
metric~\eqref{eq:nab-metric} have different scales for $t$ and
$\vec{x}$ (although see~\cite{George:2008vu} for a possible
way of dealing with this problem).

To proceed we therefore consider long wavelength perturbations
such that $\vec{x}$ derivatives of the perturbations are set to
zero, and hence $g_{AB}$ \eqref{eq:deltaG} is independent of the
3-spatial coordinates.  The gauge freedom is no longer fixed
completely.  Performing a gauge transformation $\delta x^M(t,y)$
with $x^M = \{t,x,y\}$ leaves the metric invariant.  This is because
$B,B_y,E$ will change by a $(t,y)$ dependent function, but since they
all appear with spatial derivatives in the metric, the metric remains
invariant.  We use this gauge freedom to set $J=0$ and $F=G=H/2$,
which leaves us in the same gauge used for the pure soft-wall case.

Consider then the metric
\be
\dd s^2 = \e^{-2\sigma(z)} \left\{ -f(z) [1-2F(t,z)] \dd t^2
+ [1-2F(t,z)] \dd \vec{x}^2 
+ f(z) [1+4F(t,z)] \dd z^2 \right\} \:,
\ee
where the $z$-coordinate is chosen (as opposed to the original $y$
frame) such that the resulting equation of motion for the
physical mode is of the Schr\"odinger form.  Further, we perturb
the scalar $\phi(t,z) = \phi(z) + \delta \phi(t,z)$.  The
equations of motion are solved for
\be
\kappa^2 \phi' \delta \phi =
  F' - 2\sigma' F - \frac{f'}{2f} F \:,
\ee
which can be used to eliminate $\delta\phi$.  Note that a prime in
this section denotes derivative with respect to $z$.  Rescaling
$F$ via
\be
F = \frac{\kappa\phi'}{k} \e^{\frac32\sigma} \tilde F \;,
\label{eq:tildeFdefn}
\ee
the system consists of a first order constraint equation and a
Schr\"odinger equation, which are, respectively,
\begin{align}
\frac{f'}{f} \left( \phi' \, \e^{\frac32\sigma} \tilde F \right)' &= 0 \;,
\label{eq:constrF}\\
\ddot{\tilde F} - {\tilde F}''+ V_F \tilde F &= 0 \:,
\label{eq:schroF}
\end{align}
where the potential is
\be
V_F =
  \frac{9}{4} \sigma'^2
  + \frac{5}{2} \sigma''
  - \frac{\sigma' \phi''}{\phi'}
  + \frac{2 (\phi'')^2}{(\phi')^2}
  - \frac{\phi'''}{\phi'}
  + \frac{3 \sigma' f'}{f}
  - \frac{\phi'' f'}{\phi' f} \:.
\label{eq:VF}
\ee
The zeroth order boundary conditions at the brane position are
\eqref{eq:BC}; at first order we find in addition
\be
(\rho_\text{br} + p_\text{br}) {\tilde F}(0) = 0 \:.
\label{eq:BCF}
\ee

For $f'=0$ we recover the soft-wall background, and the above
equations yield the correct result: the constraint
equation~\eqref{eq:constrF} is satisfied, $V_F$ is equivalent to
the soft-wall form~\cite{Cabrer:2009we}, and the boundary
condition~\eqref{eq:BCF} is satisfied since
$(\rho_\text{br}+p_\text{br})=0$.

When $f'\ne0$ we have a black-wall background, with event
horizon at $z=z_h$ defined by $f(z_h)=0$.
At first sight it may seem that equation~\eqref{eq:constrF}
requires the solution
\be
{\tilde F} = F_0 k \frac{\e^{-\frac32\sigma}}{\kappa\phi'} \:.
\ee
The boundary condition~\eqref{eq:BCF} then forces $F_0=0$ hence
$\tilde F=0$ everywhere.  Thus all long wavelength perturbations
vanish.  But this conclusion is only valid when $f'/f$ is much
larger in magnitude than the perturbation $F$.

The general solution for $f$ is given by equation~\eqref{eq:solh}
and, as previously discussed, $A$ should be kept very small so
that the horizon is as close as possible to the singularity.  Then
it should be that the generic features of the soft-wall (for
example the KK spectra) are retained.  For perturbations that have
magnitude of order, or larger than, $A$, the factor
$f'/kf=A\e^{3\sigma}$ is now perturbatively small in the region
leading up to the horizon.  Since the derivation of the
perturbation equations for ${\tilde F}$ are only valid to first
order, we should therefore not include higher order terms such as
$(f'/kf){\tilde F}$.  Furthermore, the ghost matter contribution
$(\kappa^2/k)(\rho_\text{br}+p_\text{br})$ is also order $A$ and
thus perturbatively small.

This means, for a given small $A$, there are three regions in $z$
to consider.  The first is where the black-wall contributes only
perturbatively, so that $A\e^{3\sigma(z)} \lesssim {\tilde F}$.
The second region is where the black-wall background starts to
dominate the solution, and the third region has the black-wall
completely washing out the perturbation ${\tilde F}$, defined by
$A\e^{3\sigma(z)} \gg {\tilde F}$.  Note that these regions always
exist, no matter how small $A$ is.

In the first region equation~\eqref{eq:constrF} is second order
(or greater) in ${\tilde F}$ and so is identically zero to first 
order in perturbations (the order to which we are working).
Similarly, the boundary condition~\eqref{eq:BCF} is second order
and so provides no constraints.  In the potential $V_F$ the last
two terms contain the factor $f'/f$ and are subdominant compared
with the other terms, hence should be discarded.  In this first
region, the system therefore reduces to the usual soft-wall case.
This is as expected since $f\sim1$.

The second region is a transition from normal soft-wall behaviour
to the dominance of the black-wall horizon, and serves simply to
provide a matching of the effective Schr\"odinger equations valid
at the two extremes.  The KK spectrum is not expected to depend 
critically on the middle region.

In $z$ coordinates, the horizon is approached as $z\to\infty$ (it
must be infinite, otherwise the black-wall is not black).  Thus
the third region has the domain $z\in[z_3,\infty)$ for some large
but finite $z_3$.  For $z$ in this domain, the asymptotic
behaviour of the background functions is found to be
\begin{align}
f(z) &\to \e^{-2m} \, \e^{-2lz} \left( 1 + \frac{n}{l} \e^{-2lz} \right) \:,\\
\sigma(z) &\to \frac13 \ln\left(\frac{-2l}{Ak} \right) + \frac{n}{3l} \e^{-2lz} \:,\\
\phi(z) &\to \phi_H + \frac{n}{3l\kappa} \e^{-2lz} \:,
\end{align}
where $l,m,n$ are free constants which parameterise the asymptotic
behaviour.  The horizon value of the scalar, $\phi_H$, is defined
by
\be
V(\phi_H) = \frac{2ln}{3\kappa^2} \e^{2m} \left(\frac{-2l}{Ak}\right)^{\frac23} \:.
\ee
Given this behaviour for the background we can determine the
allowed solutions for the perturbation ${\tilde F}$.  The boundary
condition at the horizon is $F(y_h)=0$, which is satisfied for
finite ${\tilde F}$ at the horizon due to the $\phi'\sim\e^{-2lz}$
factor in equation~\eqref{eq:tildeFdefn}.\footnote{$F(y_h)=0$ only
requires ${\tilde F}$ to diverge slower than $\e^{-2lz}$.  But, as
usual, we require ${\tilde F}$ to be finite so that it is 
plane-wave normalisable at infinity.}   Similarly, one can deduce
that $\partial_y F(y_h)$ is finite, $\partial_z F\to0$, but
$\partial_z {\tilde F}$ is again allowed to be finite at the
horizon.  As for the Schr\"odinger equation for ${\tilde F}$, the
asymptotic form of $V_F$ is
\be
V_F \to 2ln \e^{-2lz} \:,
\ee
which goes to zero.  Combining this with the above boundary
conditions we find that the KK spectrum in region three is simply
plane waves.  Going back to physical $y$ coordinates, these
plane waves become squeezed into a tiny finite region just before
the horizon at $y_h$.

To get the full KK spectrum of the black wall we must match region
one, the usual soft-wall spectrum, with the plane waves near the
horizon.  For a KK energy/mass on resonant with a soft-wall mode,
the wavefunction is exponentially small as it enters region three,
and the amplitude of the plane wave is almost zero.  For
off-resonant KK energies, the plane wave has sizeable amplitude
and after the wavefunction is properly normalised (which happens
in the $z$ coordinate) it has almost zero amplitude in the first
region.  Thus the black-wall KK spectrum is essentially the same
as the soft-wall spectrum, but with an overlay of continuum modes
starting from zero energy which are strongly localised toward the
horizon.  Such behaviour is reminiscent of RS2 domain-wall
models~\cite{Davies:2007tq}.


\section{Unitary boundary conditions}
\label{sec:BC}

As seen in Section~\ref{sec:bw}, shielding the singularity by a
horizon requires ghost matter.  But such a construction may not be
necessary.  We shall argue in this section that the naked
singularity obeys unitary boundary conditions which assures no
conserved charges can leak away~\cite{GellMann:1985if,
Cohen:1999ia, Gremm:2000dj}.  Consequently, unitarity is conserved
in the theory and a sensible
bulk physics can be defined. In a way, the singularity is not
``visible'', and thus no shielding by a horizon is needed.

The soft-wall singularity is located at a finite coordinate distance
in the phenomenologically interesting parameter range $\nu  \geq 1$.
Naively one might think this is problematic, as light can travel to it
in a finite time.  The time it takes light to travel to the boundary
follows from the metric \eqref{eq:metric}:
\be
\Delta t= 
\int_{0}^{y_s} \frac{\e^{\sigma}}{f} \dd y 
= \left\{
\begin{array}{cc}
\text{(finite)}
  + \int^\infty
    (\kappa\phi)^{-\beta(1+1/\nu^2)} \;
    \e^{-\nu\kappa\phi(1-1/\nu^2)} \;
    \dd \phi,  \quad & (f=1),\\
\infty, &  (f\neq 1).
\end{array}
\right.
\ee
For the soft-wall case with $f=1$, we have split the integral into
the bulk region and the region near the singularity.  The bulk always
gives a finite contribution, and we can thus focus on the singularity
region only.  As $y \to y_s$ the field diverges $\phi \to \infty$ and
the superpotential is of the asymptotic form \eqref{eq:limW}.  The
integral is finite for $\nu=1$ and $\beta >1/2$, and also for $\nu >1$
with arbitrary $\beta$.  For the black-wall case, $f\ne1$, we must
have an infinite time-interval, otherwise light can escape from the
horizon, meaning it is not actually black.  Thus $\e^\sigma/f$ must
diverge at least as fast as $(y_h-y)^{-1}$ at the horizon.

The key point, however, is that there are no modes in the spectrum
that can travel freely along the bulk direction.  The KK spectrum
of the soft-wall solution is discrete (for $\nu=1$ continuous, but
the same reasoning applies), with all bound states satisfying the
boundary conditions (\ref{eq:surface1},~\ref{eq:surface2}) at the
singularity.  All perturbative modes have an extra dimensional profile
that dies off rapidly at the singularity, and consequently these modes
cannot really probe the boundary.  As we shall see, this behaviour
means that no 4D energy or momentum can leak into the singularity.

The energy-momentum current density is $J^M = T^{MN} \xi^{(\mu)}_N$
with $T^{MN}$ the energy-momentum tensor \eqref{eq:T}, and $\xi^{(\mu)}_N
= \delta^\mu_N$ the Killing vector generating 4D translations.
The labelling is such that the 4D index $\mu$ corresponds to current
of energy ($\mu=0$) and spatial momentum ($\mu = 1,2,3$) respectively;
the 5D index $M$ labels the direction of the current.  The current
satisfies the conservation law
$\partial_M (\sqrt{g} J^M)/\sqrt{g} =0$, which expresses
conservation of 4D energy and momentum.  To ensure that energy and
momentum remain conserved in the presence of a singularity, we demand
that the flux into the singularity vanishes
\be
\sqrt{g}J^y|_{y_s} = 0.
\ee
For the tensor perturbations we can use the result of
\cite{Cohen:1999ia, Gremm:2000dj}
\be
\sqrt{g}J^y|_{y_s} = \sqrt{g}g^{yy} \frac12 h_{kl}' 
\partial_i h_{kl}|_{y_s}
\propto \e^{-4\sigma} hh'|_{y_s} =0
\ee
where in the second step we used separation of variables
\eqref{eq:separation}. This vanishes exactly for all KK modes
because of the boundary condition \eqref{eq:surface1}.  
Expanding the energy-momentum tensor in the scalar perturbations
defined in Section \ref{sec:KK} gives
\ba \sqrt{g}J^y|_{y_s} &=&
\e^{-4 \sigma}\[ \phi'  
+ ( \delta \phi'-10 F \phi') \]\partial_i \delta \phi
\big|_{y_s} \nn \\
&\propto& 
\e^{-4 \sigma}\[ \phi' \delta \phi  
+ \( \delta \phi \delta \phi'-  \frac{10}{\kappa^2}(F F' -2\sigma' F^2) \)\] 
\bigg|_{y_s} 
\ea
where in the second line we again used separation of variables
\eqref{eq:separation}. Moreover, we used that near the singularity
$\phi' = \sigma'$, and also the constraint equation relating $F$
and $\delta \phi$ \eqref{eq:constraintF}.  The terms first and
second order in the perturbations all vanish independently due to
the boundary conditions \eqref{eq:surface2}, and hence this
flux vanishes at the singularity.

We conclude that, at least in the perturbative regime, no 4D
energy or momentum leaks into the singularity, and there should be
no problems with loss of unitarity.


\section{Conclusions}
\label{sec:concl}

We have carried out a critical analysis of the singularity in
soft-wall models.  The general idea is to make sure that this
singularity is not ``visible'', to protect low energy physics from
uncontrollable phenomena such as non-unitarity, loss of conserved
quantities, or unknown quantum gravity.  

In the soft-wall literature to date it has been assumed that a
black-wall horizon is needed to hide the singularity.  With this
in mind, we first set-up the general equations for a black-wall.
Working in the fake supergravity formalism, we assumed a
superpotential of the asymptotic form
\be
\lim_{\phi \to \infty} W \propto c (\kappa\phi)^\beta \e^{\nu\kappa\phi} \:,
\ee
A phenomenologically viable soft-wall solution solving the hierarchy
problem can be constructed for $c > 0$ and $1 \leq \nu < 2$.  In the
bulk, this solution can be extended to a black-wall solution with a
horizon shielding the singularity.  The KK spectrum of such a set-up
is qualitatively similar to the soft-wall case, as discussed in
Section~\ref{sec:bwpert}.  However, matching the Israel junction
conditions at the origin requires ghost matter on the UV brane.
Even though the density of this ghost matter can be made tiny, the
set-up is rather unsatisfying.  To avoid ghost matter one could
try to look for a time-dependent but cosmologically long-lived
black-wall solutions, although the naive expectation is a decay
time of order the electroweak scale.

However, as we further pointed out in this paper, a horizon is not
in fact necessary as the singularity is not ``visible'' to any of
the states in the spectrum.  Indeed, the extra-dimensional profile
of all KK modes decay rapidly near the singularity; this is needed
to obtain a vanishing 4D cosmological constant, and thus a
phenomenologically viable model.  This decay also ensures that the
system satisfies unitary boundary conditions.  No 4D energy or any
other conserved quantity can leak into the singularity, and
consequently a well-defined unitary theory can be constructed.


\acknowledgments
This research was supported by the Netherlands Foundation for
Fundamental Research of Matter (FOM) and the Netherlands
Organisation for Scientific Research (NWO).



\providecommand{\href}[2]{#2}\begingroup\raggedright\endgroup

\end{document}